\documentclass[prl,twocolumn,showpacs,]{revtex4}
\usepackage{ae}
\usepackage{graphicx}
\usepackage{amssymb}
\usepackage{amsmath}
\usepackage{amsthm}
\usepackage{color}
\usepackage{sidecap}
\usepackage{ulem}
\usepackage{wasysym}
\usepackage{pifont}
\usepackage{ifsym}
\def\dg{\dagger}
\def\up{\uparrow}
\def\dn{\downarrow}
\def\Uu{U_{\up\up}}

\def\Uud{U_{\up\dn}}
\def\Ju{J_{\up}}
\def\Jd{J_{\dn}}
\def\Nup{N_{\up}}
\def\Ndn{N_{\dn}}
\def\uphi{U_{\phi}}

\newcommand{\bdn}[1]{\hat{b}_{#1\dn}}
\newcommand{\bdagdn}[1]{\hat{b}_{#1\dn}^{\dagger}}

\newcommand{\OP}[2]{\hat{#1}_{#2}}

\begin{document}
\title{Competing regimes of motion in 1D mobile impurities}

\author{A.~Kantian$^1$, U.~Schollw\"ock$^2$, T.~Giamarchi$^1$}
\affiliation{(1) DPMC-MaNEP, University of Geneva, 24 Quai Ernest-Ansermet, 1211 Geneva, Switzerland\\
  (2) Department f\"ur Physik, LMU M\"unchen, Theresienstrasse 37, 80333 M\"unchen, Germany }

\begin{abstract}

We show that a distinguishable mobile impurity inside a one-dimensional many-body state at zero
temperature generally does not behave like a quasiparticle (QP). Instead, both the impurities
dynamics as well as the ground state of the bath are fundamentally transformed
by a diverging number of zero-energy excitations being generated, leading to what we call infrared-dominated (ID) dynamics.
Combining analytics and DMRG numerics we provide a general formula for the power law governing ID dynamics
at zero momentum, discuss a threshold
beyond which quasiparticle dynamics may occur again,
and study the competition between the ID and quasiparticle universality classes at larger impurity momenta.
\end{abstract}


\date{\today}

\maketitle

The dominant paradigm to describe the motion of a single, distinguishable quantum particle
inside quantum liquids has been that of the \textit{quasiparticle} (QP). It was originally motivated
by the study of an electron moving in a bath of phonons, i.e. the polaron. In this model, the impurities' dynamics
remain coherently ballistic, albeit now with a finite lifetime, and its mass is renormalized.
Simultaneously, the impurities' effect on the bath is limited to a local correlation hole forming around it
- otherwise, impurity and bath behave independently. 
The quasiparticle concept has seen very widespread use. Beyond polarons, it is employed for
impurities in various Fermi liquids~\cite{Kopp1990,Rosch1995}, and has been extensively applied
to experiments on ultracold gases~\cite{Schirotzek2009,Nascimbene2009,Koschorreck2012,Kohstall2012}.
Recently, experiments have focused on impurities moving specifically inside one-dimensional
gases~\cite{Palzer2009,Catani2012,Fukuhara2013,Fukuhara2013a}.

A question of clear interest is then whether the quasiparticle is actually the only class of dynamics
describing impurity motion inside every quantum liquid. There were indications to the contrary.
The experiments on cold atom gases in 1D had in part been stimulated by theory showing that
specific one-dimensional integrable models, marked by special parameter values for impurity and bath,
exhibit a complete breakdown of QP physics~\cite{Castella1993,Zvonarev2007,Zvonarev2009,Zvonarev2009a}.
The solutions for these special models show not only the loss of the relationship between impurity energy and momentum
central to QP behaviour, but moreover that impurity correlations spread
subdiffusively - i.e. logarithmically slow - within a window of time~\cite{Zvonarev2007}, very different from the ballistic propagation
of the QP. For a free fermion bath, the mechanism behind this drastic change of dynamics in 1D was
described as onset of an \textit{Orthogonality Catastrophe}~\cite{Kopp1990,Rosch1995}, a diverging number of low-energy excitations emitted by the impurity into the bath, leading to a reordering of its ground state~\cite{BookMahan2000}. 
We call this the \textit{infrared-dominated} (ID) regime in the following. 

It was conjectured that ID-dynamics of impurities in one-dimensional baths, driven by the Orthogonality Catastrophe, could be universal, in  contrast to the naive
expectation that the QP model applies in all dimensions. Despite work on a general theory~\cite{Zvonarev2009,Lamacraft2009,Kamenev2009} in the limit of small impurity momenta - using a Tomonaga Luttinger liquid-based (TLL)~\cite{BookGiamarchi2003} field theory description - the conjecture had remained open. Specifically, it was unknown whether those predictions of universal ID
physics hinged on linearizing the baths spectrum, up to which impurity momentum ID physics actually persists,
and whether there were competing dynamics. Finally, from this previous work the quantities characterizing ID physics generally could only be obtained implicitly.

In this article, we show that
ID dynamics indeed are universal for mobile impurities in 1D, using analytic and Density Matrix renormalization group
(DMRG)~\cite{Schollwock2011} techniques. We provide a sufficient criterion for the establishment of ID behaviour and show that QP
dynamics may occur when it is violated, depending subtly on the interactions and the baths excitation spectrum. We also provide 
an explicit description of ID dynamics for any system at zero momentum, using only static properties.

The standard way to describe the propagation of a distinguishable impurity - denoted
by a $\dn$-label - over distance $x$ during time $t$ inserted inside a bath of $\up$-particles 
is through the Green's function
\begin{equation}\label{def:gf}
G(x,t) = -i\langle \hat{b}_{x\downarrow}(t) \hat{b}^{\dagger}_{0\downarrow}(0) \rangle,\quad t>0
\end{equation}
where $\bdagdn{x}$ ($\bdn{x}$) denotes the impurities' creator/annihilator, and translational invariance
allows Fourier transformations into
$G(p,t)$ and $G(p,\omega)$. 
\begin{figure}[t]
\includegraphics[width=1\columnwidth,trim = 50mm 12mm 40mm 0mm, clip]{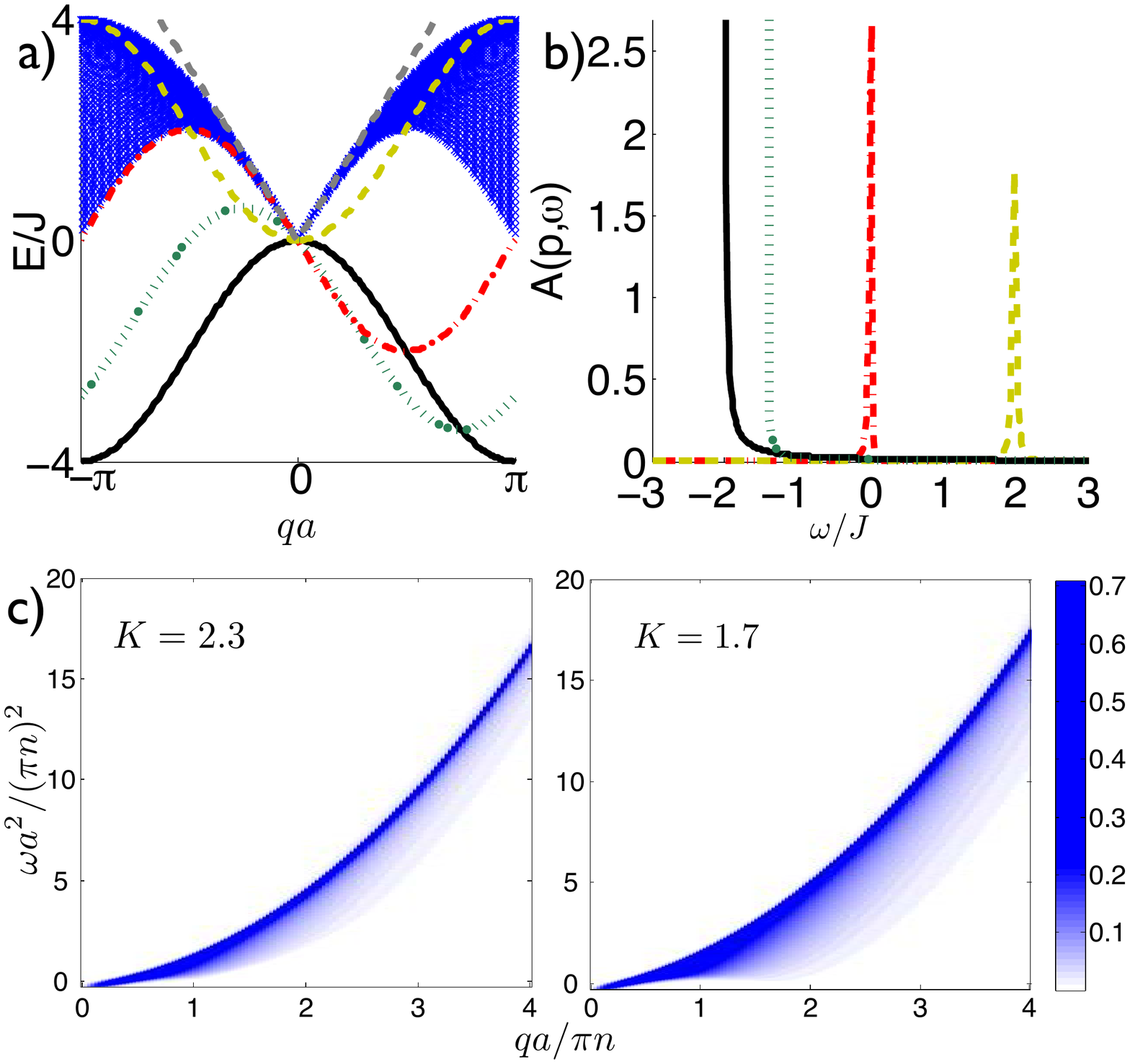}
\caption{\protect\textbf{a)} Single-particle excitation spectrum (SPES) for free 1D lattice-fermions, with $k_F=\pi/2a$ (shaded area). Low-momentum linearization $v|q|$
also shown (grey dashed line), and $\delta\varepsilon(q)=\varepsilon_p-\varepsilon_{p+q}$ for $p=0$ (black, solid),
$p=\pi/4a$ (green, dotted), $p=\pi/2a$ (red, dash-dotted) and  $p=\pi/a$ (yellow, dashed). \protect\textbf{b)} Spectral function $A(p,\omega)$ of the mobile impurity, calculated with
Linked Cluster expansion, with values of $p$ and linestyles matching a). When $\delta\varepsilon(q)$ intersects with the SPES
only for $q=0$ (e.g. at $p=0$, $\pi /4a$), the impurity loses all QP properties and enters the ID regime. When the impurity can generate real phonons
in the bath, QP behaviour is restored. \protect\textbf{c)} SPES of weakly repulsive bosons in the 1D continuum, (data by J. S. Caux, c.f.~\protect\cite{Caux2006}), with TLL parameters $K=2.3$ and $K=1.7$, corresponding, at small $q$, to DMRG simulations with
$\Nup/L=0.5$, $\Uu=2\Ju$ and  $\Uu=4\Ju$ respectively.
}
\label{fig:Fig1}
\end{figure}

In the QP picture, the bath renormalizes the impurities' properties, like its dispersion relation $\epsilon_p$.
Its spectral function $A(p,\omega) := -\frac{1}{\pi} \operatorname{Im}[G(p,\omega)] $
broadens from a $\delta$-peak to $A(p,\omega)\propto \frac{1}{(\omega-\epsilon_p)^2+\tau^{-2}}$,
with $\tau$ the lifetime of coherent propagation.

As we show, inside 1D baths the impurity dynamics
is generally fundamentally different. Its spectral function changes to~\cite{Zvonarev2007,Akhanjee2007,Kamenev2009}
\begin{equation}\label{subdiff_spec_func}
A(p,\omega)  \propto \frac{\theta(\omega-\epsilon_{p})}{(\omega-\epsilon_{p})^{\Delta(p)}}
\end{equation}
for $\omega$ close to $\epsilon_p$,  whose sole remaining
role is to determine the domain of $A(p,\omega)$ ($\theta(\omega)$: step-function), as opposed to the 
QP regime, where $\epsilon_p$ still defines an energy-momentum relationship.
The algebraic power $\Delta(p)$ is momentum-dependent and can be expanded as 
$\Delta(p)\approx \Delta(0)+\beta p^2$.

\begin{figure}[t]
\begin{center}
\vspace{-0cm}
\includegraphics[width=1\columnwidth,trim = 88mm 28mm 88mm 62mm, clip]{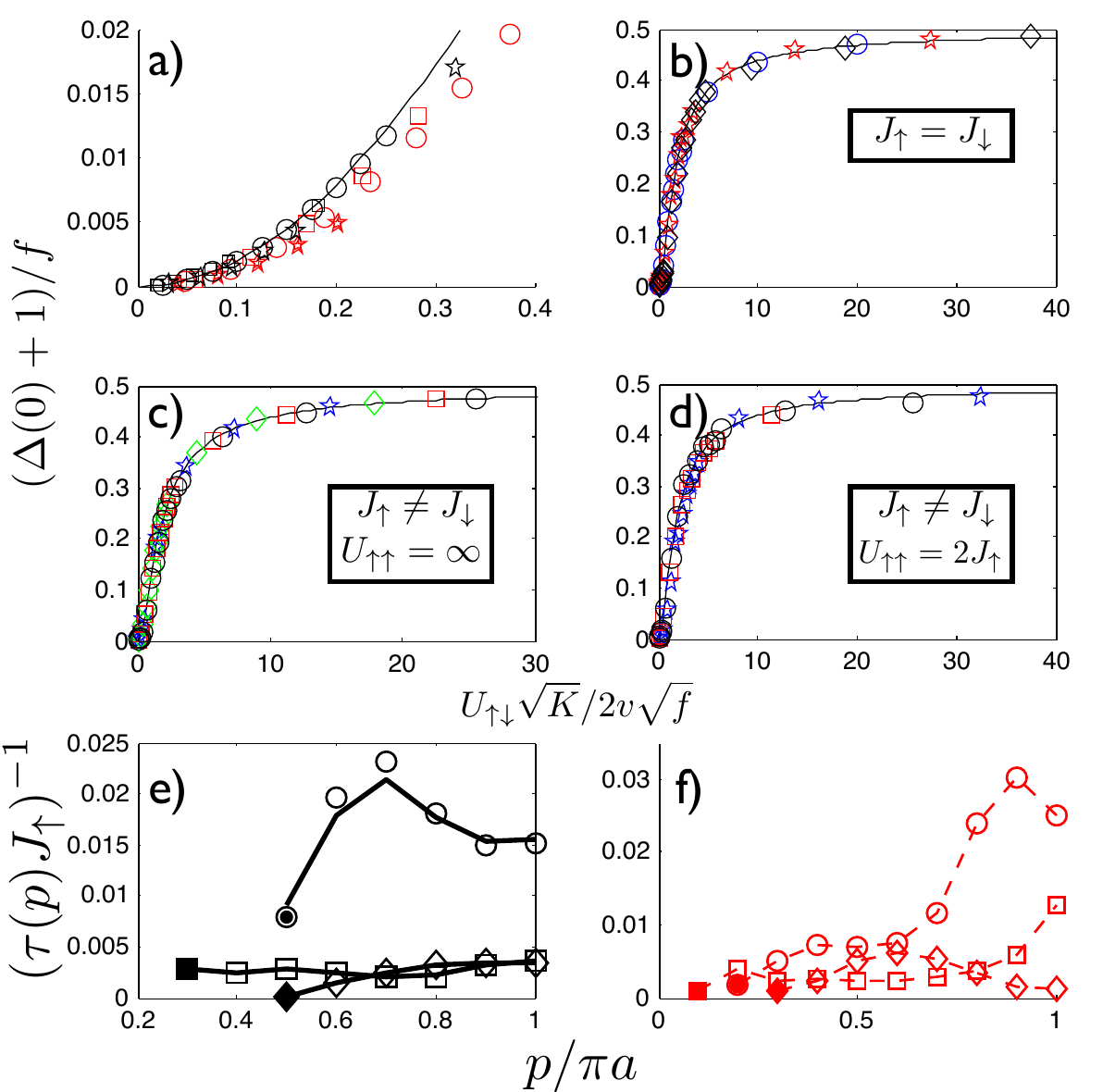}
\end{center}
\caption{ \textbf{a)} Data collapse of DMRG results (markers) and $\Delta_{\rm LCE}(0)$ (black line) at $\Uud=0.2\Ju$ showing the agreement between DMRG and LCE.  Black
markers: free fermions, $\Uu=\infty$; red markers: softcore bosons, $\Uu=2\Ju$. $\Jd=0.5\Ju$ ($\Square$),
$\Jd=\Ju$ ($\circ$), $\Jd=2\Ju$ (\ding{73}).
\textbf{b) - d)} Data collapse of DMRG results and $\Delta(0)$ from eq. (\ref{delta_analytical}) (black line).
\textbf{b)} $\Ju=\Jd$, $\Uu=\infty$ ($\circ$), $\Uu=4J$ (\ding{73}), $\Uu=2J$ ($\Diamond$).
\textbf{c)} Free fermions, $\Uu=\infty$, $\Jd=0.5\Ju$ ($\circ$), $\Jd=0.75\Ju$ ($\Square$), $\Jd=1.33\Ju$ ($\Diamond$),
$\Jd=2\Ju$ (\ding{73}). \textbf{d)} Softcore bosons, $\Uu=2\Ju$, $\Jd=0.2\Ju$ ($\circ$), $\Jd=0.5\Ju$ ($\Square$),
$\Jd=2\Ju$ (\ding{73}).
\textbf{e) - f)} Inverse QP lifetime $\tau(p)^{-1}$ vs. $p$, at $\Uud=0.2\Ju$,
for $\Jd=\Ju$ ($\circ$), $\Jd=0.5\Ju$ ($\Diamond$), $\Jd=2\Ju$ ($\Square$),
Lifetimes can only be defined above a certain
value of $p$, indicated by full marker, as dynamics are always ID when $p$ is small, (see main text).
\textbf{e)} Free fermions, $\Uu=\infty$, with $k_F=\pi/2a$. Lines are calculated with LCE. 
\textbf{f)} Softcore bosons, $\Uu=2\Ju$, lines are guide to the eye.
}
\label{fig:Fig2}
\end{figure}

In time and space, (\ref{subdiff_spec_func}) corresponds
to correlations spreading slower than any power law - i.e. subdiffusively - scaling as $\log(t)$ within
a parameter-dependent window of time,
in contrast to the regular (quasi-)particle behaviour, where correlations propagate linearly
with $t$~\cite{Zvonarev2007}. This subdiffusive behaviour of $G(x,t)$ translates to power law decay
of $G(p,t)$ in time~\cite{Zvonarev2007}, which we focused on.
To represent the impurity moving inside some 1D quantum liquid, we consider a generic
two-species Bose-Hubbard Hamiltonian on a chain of length $L$ governing time evolution in eq.~(\ref{def:gf}),
\begin{equation}\label{def:ham}
\hat{H} = -\sum_{\langle i, j\rangle,\sigma}J_{\sigma}\hat{b}^{\dagger}_{i\sigma} \hat{b}_{j\sigma}+\sum_{i,\sigma,\sigma'}\frac{U_{\sigma\sigma'}}{1+\delta_{\sigma\sigma'}}\hat{n}_{i\sigma}(\hat{n}_{i\sigma'}-\delta_{\sigma,\sigma'}),
\end{equation}
where $\sigma=\uparrow,\downarrow$. The ground state of $\hat{H}$ used in (\ref{def:gf}) is that with $\Nup=N$, $\Ndn=0$ at incommensurate filling (thus, a Mott-insulating state cannot occur), with quantum-liquid characteristics adjustable by tuning  $\Ju$, $\Uu$ and $\Nup$.
Interactions in (\ref{def:ham}) are assumed to be repulsive.
For an analytical understanding of the impurity dynamics, we employ the
\textit{Linked Cluster Expansion} (LCE)~\cite{BookMahan2000} to quadratic order in $\Uud$ 
to compute $G(p,t)$ approximatively. Results from DMRG
show this method working very well for small $\Uud$ everywhere we checked,
as exemplified in Figs.~\ref{fig:Fig2}a and e, obviating the need to continue LCE to third order in $\Uud$.

Applying LCE to (\ref{def:ham}) requires linearizing
its low-energy excitation spectrum using the 
Tomonaga-Luttinger liquid (TLL) approximation (always possible \footnote{See Supplemental Material for details}). Its properties
are determined by the velocity of bath excitations $v$, the
parameter $K$ (both can be obtained from DMRG), and a cutoff momentum $q_c$.
We then obtain
\begin{equation}\label{lce1}
G_{\rm LCE}(p,t) = -i e^{-i\varepsilon_p t}e^{F_2(p,t)}
\end{equation}
where $\varepsilon_p = -2\Jd\cos p$ and $F_2(p,t)=-\int du \frac{1+itu-e^{itu}}{u^2}R(u)$~[23]. Here $R(u)=\frac{K\Uud^2}{2\pi}\int dq |q|e^{-|q|/q_c}\delta(u+\varepsilon_p-\varepsilon_{p+q}-v|q|)$, is the density
of impurity-generated excitations. As we focus on $G(p,t)$ at long times,
the properties of $R(u)$ at small $u$ are crucial. We find that $R(u)\propto u$
at small $u$, as long as $\delta\varepsilon(q):=\varepsilon_p-\varepsilon_{p+q}$
intersects with the phonon dispersion $v|q|$ only at $q=0$ (c.f. Fig.~\ref{fig:Fig1}a, c) - the impurity only causes the generation of 
virtual phonons. This is the signature behaviour of $R(u)$ observed when describing the occurrence
of the Orthogonality Catastrophe for the immobile impurity in Fermi liquids using LCE~\cite{BookMahan2000}, where low-energy single-particle excitations
dominate the dynamics too, changing them from QP to ID-dominated. Conversely, when 
$\delta\varepsilon(q)=v|q|$ holds for $q\neq0$, and thus the impurity emits real phonons, $R(u)$ is near-constant for small
$u$, low-energy excitations no longer dominate, and $F_2(p,t)\propto t$, restoring QP behaviour. 
When $\varepsilon_p\approx\Jd p^2$ is valid, this criterion translates to the impurity having ID-dynamics when its
'velocity' is below the baths sound velocity, $2\Jd p<v$, with $\Delta_{\rm LCE}(p)=-1+\frac{K\Uud^2}{2\pi^2v^2}\left(1+\frac{6\Jd^2}{v^2} p^2\right)$, and being QP-like when $2\Jd p>v$~[23].

Taking the 
bath's full single-particle excitation spectrum (SPES) into account, this \textit{sufficiency criterion} separating ID from QP dynamics 
is modified. When the $Ô\upÔ$-particles are free fermions, $R(u)$ is known exactly~[23]:
\begin{equation}\label{lce2}
R(u) = \int_{\stackrel{|k_1|>k_F}{|k_2|<k_F}}  dk_1 dk_2 \delta(u + \delta\varepsilon(k_1-k_2) + \omega_{k_1} - \omega_{k_2}),
\end{equation}
where $\omega_k = -2\Ju\cos k$ is the dispersion
relation of the $\up$-particles. Calculating the spectral function $A_{\rm LCE}(p,\omega)$ 
(c.f. Fig.~\ref{fig:Fig1}b) shows that when $\delta\varepsilon(q)$ intersects with the full SPES only at $q=k_1 - k_2=0$, the impurity dynamics is described by eq. 
(\ref{subdiff_spec_func}), whereas an intersection at $q\neq0$ restores QP coherence. Figures~\ref{fig:Fig2}e-f provide confirmation from DMRG:
when $\Uu=\infty$, even an impurity tunneling $\Jd$ so weak that $\delta\varepsilon(q)\neq v|q|$ at all $q$ still leads to
ID-regime breakdown and reemergence of the quasiparticle one for $p\geq k_F$,  as shown in Fig.~\ref{fig:Fig2}e, the full SPES bending to zero again at $q=2k_F$ with finite weight (Fig.~\ref{fig:Fig1}a). This makes $p=k_F$ the 'default' momentum above which ID dynamics break down for free fermions - unless $\Jd>\Ju$, when this occurs earlier due to intersection of $\delta\varepsilon(q)$ with the SPES at small $q$ (c.f. Fig.~\ref{fig:Fig2}e).
Comparison with the opposite limit of softcore bosons ($\Uu=2\Ju$) is instructive: there, the momentum above which ID behaviour 
is replaced by a QP one is only governed by the intersection of $\delta\varepsilon(q)$ with the phonon spectrum around $q=0$, (Fig.~\ref{fig:Fig2}f). As
depicted, quasiparticle lifetimes start being defined at successively higher $p$ the lower $\Jd$ becomes, due to the SPES
of softcore bosons having very low weight at low energies, except at $q\sim0$ (Fig. \ref{fig:Fig1} c). Thus, for sufficiently small $\Jd$,
it can happen, as we observe e.g. at $\Jd=0.2\Ju$~[23] that DMRG shows ID dynamics at \textit{any} $p$ within simulation times,
in contrast to a free fermion bath.
Finally, this analysis explains why the ID-regime is confined to 1D baths: only here has the SPES exponentially small weight at zero energy within a momentum range (c.f. Figs. \ref{fig:Fig1}a, c). Based on the properties of $R(u)$, we predict that in the one higher-dimensional case where the SPES has a similar structure - i.e. the near-perfect superfluid with only a Bogoliubov-like excitation branch - ID physics still does not occur due to the larger phase-space leading to $R(u)\not\propto u$.

Simulations complement the LCE, by computing
$G(p,t)$ at large $\Uud$, and validate the LCE results at small $\Uud$. We calculate
$G_{\rm DMRG}(x,t)$ using t-DMRG~\cite{Schollwock2011}, by inserting the impurity into the ground state 
of (\ref{def:ham}) in the fully polarized spin sector $\Nup=N$, $\Ndn=0$
at $t=0$, in the center of a lattice with $L=201$ sites. We simulate for $\Nup\in[20,34,50,66,80,100,133]$
(i.e. close to the simple filling fractions $\frac{1}{10}$, $\frac{1}{6}$, $\frac{1}{4}$, $\frac{1}{3}$, $\frac{2}{5}$, $\frac{1}{2}$,
$\frac{2}{3}$).
Shown results are for $\Nup=100$ unless noted otherwise.
The resultant state is evolved up to time $T_f\Ju=50-70$, and the  
overlaps with states $e^{-iE_{\rm GS}t}\bdagdn{x}|GS\rangle$, $x\in[1,101]$ computed after
every step, yielding $G_{\rm DMRG}(p,t)$. When $|G_{\rm DMRG}(p,t)|$ can be fit
with a power law - while checking that
$A_{\rm DMRG}(p,\omega)$ shows threshold behaviour in accordance with~(\ref{subdiff_spec_func}) -
$\Delta(p)$ is extracted. This approach is used because DMRG simulations at long times break down due to 
exponential growth of entanglement - and high resolution in $\omega\propto T_f^{-1}$
is needed to extract $\Delta(p)$ directly
from $A_{\rm DMRG}(p,\omega)$. For different $\Uu$, this is done for bath-impurity interactions $\Uud\in[0.1\Ju,80\Ju]$.
We further examine the impact of having $\Ju\neq\Jd$, with
$\Jd\in[0.2\Ju,2\Ju]$. We frequency-filter $G_{\rm DMRG}(p,t)$,
omitting $\omega\notin[-\Uud,\Uud]$,
due to the lattice-specific appearance of additional power-law singularities for repulsively bound
states~\cite{Winkler2006} when $\Uud\geq 2\Ju+2\Jd$~[23].
\begin{figure}
\begin{center}
\vspace{-0cm}
\includegraphics[width=1\columnwidth,trim = 8mm 50mm 45mm 60mm, clip]{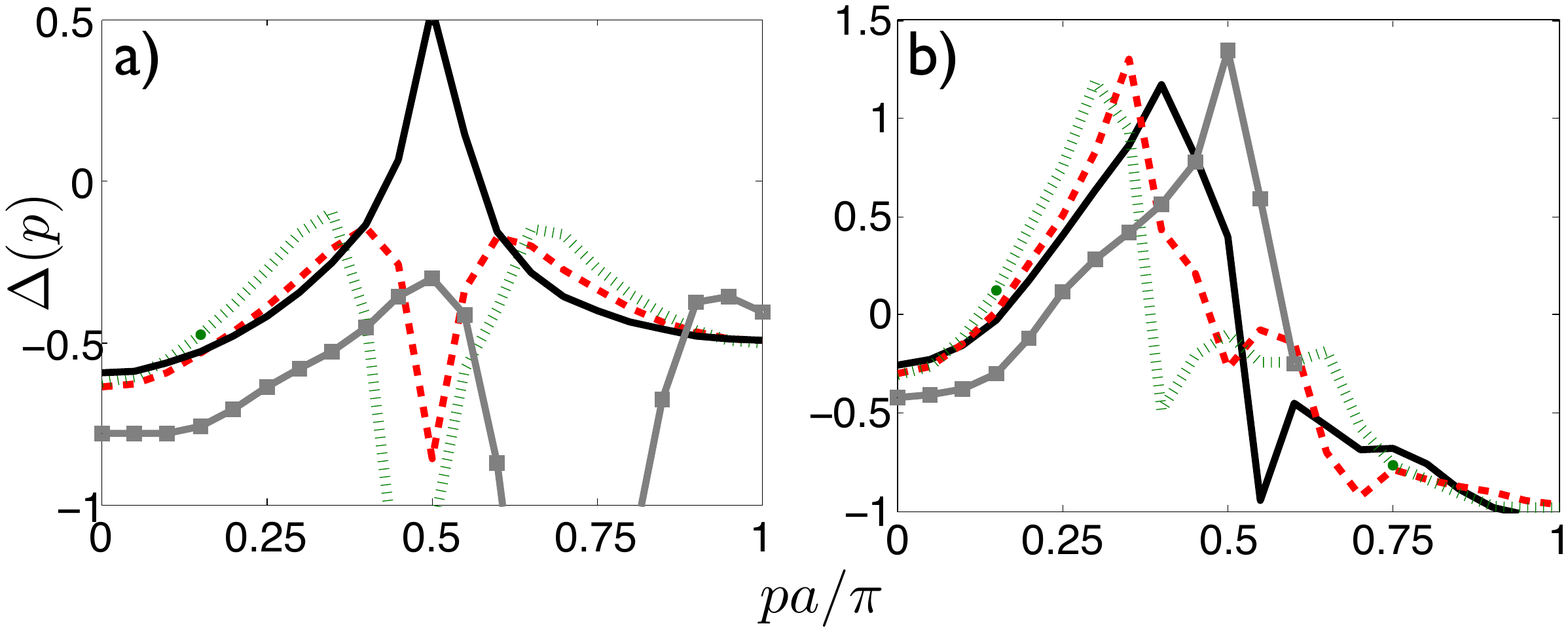}
\end{center}
\caption{Exponent $\Delta(p)$ of ID dynamics at dominant $\Uud$, calculated by DMRG for different $\Nup/L$ and $\Ju/\Jd$. Linebreaks signal that ID dynamics are displaced either by QP physics or a third, as-yet unknown regime (see main text). Black solid: $\Nup/L=0.5$, $\Jd=\Ju$; red dashed: $\Nup/L=0.4$, $\Jd=\Ju$; green dotted: $\Nup/L=0.33$, $\Jd=\Ju$; 
grey $\Square$-solid: $\Nup/L=0.5$, $\Jd=0.5\Ju$. \textbf{a)} Free fermions ($\Uu=\infty$), $\Uud=40\Ju$. For $\Jd=\Ju$ and half-filling,
the impurity is always in the ID regime. For $k_F<\pi/2a$ however, $\Delta(p)$ drops off
between $p\in[k_F,\pi/a-k_F]$. For $k_F=0.33\pi/a$, we observe a momentum range around $p=\pi/2a$, where ID dynamics are replaced by a third regime requiring further study. This regime is also observed beyond $p=k_F$ when $\Jd<\Ju$ even at $k_F=\pi/2a$ (grey $\Square$-solid).
\textbf{b)}  Softcore bosons ($\Uu=2\Ju$), $\Uud=10\Ju$. Around $p=\pi \Nup/L$, $\Delta(p)$ drops off but shows ID behaviour throughout, approaching free coherent propagation at $\pi/a$ for
$\Ju=\Jd$ (see text). Conversely, when $\Ju\neq\Jd$, QP dynamics emerges after $p$ exceeds $\pi \Nup/L$ (grey $\Square$-solid)}
\label{fig:Fig3}
\end{figure}

These simulations yield several novel results:

(i) \textit{We propose an analytical form for $\Delta(0)[\Uud]$ valid for any 1D quantum liquid, using only one additional parameter
$f$ determined by the static properties of an opaque impurity:} Data collapse (Figs.~\ref{fig:Fig2}b-d) reveals that all curves of $\Delta(0)[\Uud]$ calculated via DMRG follow
the same law. We deduce it by generalizing the exact solution for the special integrable case of free
1D fermions with $m_{\up}=m_{\dn}$~\cite{Castella1993} through the requirement that it match $\Delta_{\rm LCE}(0)$ at small $\Uud$. Making the resultant formula 
\begin{equation}\label{delta_analytical}
\Delta(0) = -1+\frac{2f}{\pi^2}\left[\arctan\left(\frac{2v}{\Uud}\sqrt{\frac{f}{K}}\right)-\frac{\pi}{2}\right]^2
\end{equation}
match the numerically obtained curves requires introduction of an additional parameter, $f$,
obtained from fitting the opaque impurity limit, $f=\lim_{\Uud\rightarrow\infty}2(\Delta(0)+1)$.
Bethe-ansatz solutions for the free fermion bath with $\Ju=\Jd$ and small-scale 
exact diagonalization for non-integrable cases suggest that for $\Uud\rightarrow\infty$ 
the ground state energy of the bath
\textit{in presence of the impurity} is $\propto$ $\Delta(0)$. It implies that low-energy, static properties of impurity and bath
determine $f$. Calculating the local density
around the impurity in its' rest frame
using the TLL approximation, then transforming to the lab frame~[23], yields
\begin{equation}\label{f_analytical}
f=\frac{L}{K}\sum_{-d\leq x\leq d}\left(\frac{\Nup}{L^2} - \langle\hat{n}_{x\up}\hat{n}_{0\dn}\rangle_{\Nup=N,\Ndn=1}\right)
\end{equation}
Here, $d$ is the extent of the correlation hole dressing the impurity in the joint ground state with the bath at
$\Uud\rightarrow\infty$.

(ii) \textit{When $\Uud$ dominates, the ID regime may
persist beyond its' limits at small $\Uud$, somewhat or substantially, depending strongly on $\Nup/L$, $\Uu/\Ju$ and $\Ju/\Jd$:}
Simulations allow to study the competition between ID and QP dynamics at large $\Uud$.
In this regime, we universally find ID dynamics persisting up to around $p=\pi\Nup/L$, with a 
$\Delta(p)$ that grows (Fig. \ref{fig:Fig3}).
Beyond this scale however, depending on $\Uu/\Ju$, $\Jd/\Ju$ and $\Nup/L$, outcomes vary strongly. For a free fermion bath with $\Ju=\Jd$ we observe 
$\Delta(p)$ dropping off for $p\in[k_F,\pi/a-k_F]$, the decay being more pronounced the lower $\Nup/L$ becomes.
At low enough $\Nup/L$, we even observe the breakdown of the ID regime within a range of $p$, and the emergence of a third, as yet unexplained regime, 
which will require further studies, indicated as linebreaks in Fig. \ref{fig:Fig3}a. When $\Jd<\Ju$, this same third regime
occurs for a range of $p>k_F$ even when $k_F=\pi/2a$ ($\Jd=0.5\Ju$ is shown in Fig. \ref{fig:Fig3}a).
However, in the reverse case $\Jd>\Ju$, once $p\geq k_F$ we find QP behaviour instead.
On the other end, for a bath of softcore bosons, the parameter $\Jd/\Ju$ has a critical influence as well. For $\Ju=\Jd$, $\Delta(p)$ drops off towards
$-1$ at $p=\pi/a$, the value for a particle scattering off a lattice-BEC of particles of equal mass (c.f.~\cite{Winkler2006}, pg. 855). As for fermions,
having $\Ju\neq\Jd$ leads to a change in regime above $p=\pi\Nup/L$, in this case from ID to QP physics ($\Jd=0.5\Ju$, 
$\Nup/L=0.5$, Fig.~\ref{fig:Fig3}b).

(iii) \textit{At small momenta, the impurity is always in the ID regime, with $\Delta(p)\approx\Delta(0)+\beta p^2$:}
In every simulation we find the ID regime persisting up to some finite momentum, with a quadratic fitting
$\alpha+\beta p^2$ to $\Delta(p)$ possible. This represents very strong evidence that ID physics is the proper paradigm to understand any mobile impurity in 1D at low momentum.

Our work has implications for future research. One is that the quasiparticle
is not the correct paradigm to understand mobile 1D impurities at low momentum. Another is that
to verify the ID regime experimentally, practically any 1D bath will suffice - provided one probes
at sufficiently low momentum. Measurements will have to be momentum resolved to avoid
convolution of signals from both regimes, except for some special cases which are always ID. Finally,
LCE analytics and time-dependent DMRG
can be relied upon for quantitative predictions of experiments.

In conclusion, we have studied a novel class of mobile impurity
dynamics in a general 1D quantum liquid, as well as the conditions for its breakdown and
the reemergence of quasiparticle behaviour.
We thank J. S. Caux for providing the data on the SPES of interacting bosons (Fig.~\ref{fig:Fig1}c). This work was supported in part by the Swiss NSF under MaNEP and division II. US acknowledges funding by the DFG through FOR801 and NIM. 
\bibliography{/Users/kantian/Documents/library.bib}

\section{The Linked Cluster Expansion to second order}
\begin{figure}[t]
\begin{center}
\vspace{-0cm}
\includegraphics[width=1\columnwidth,trim =30mm 20mm 30mm 70mm, clip]{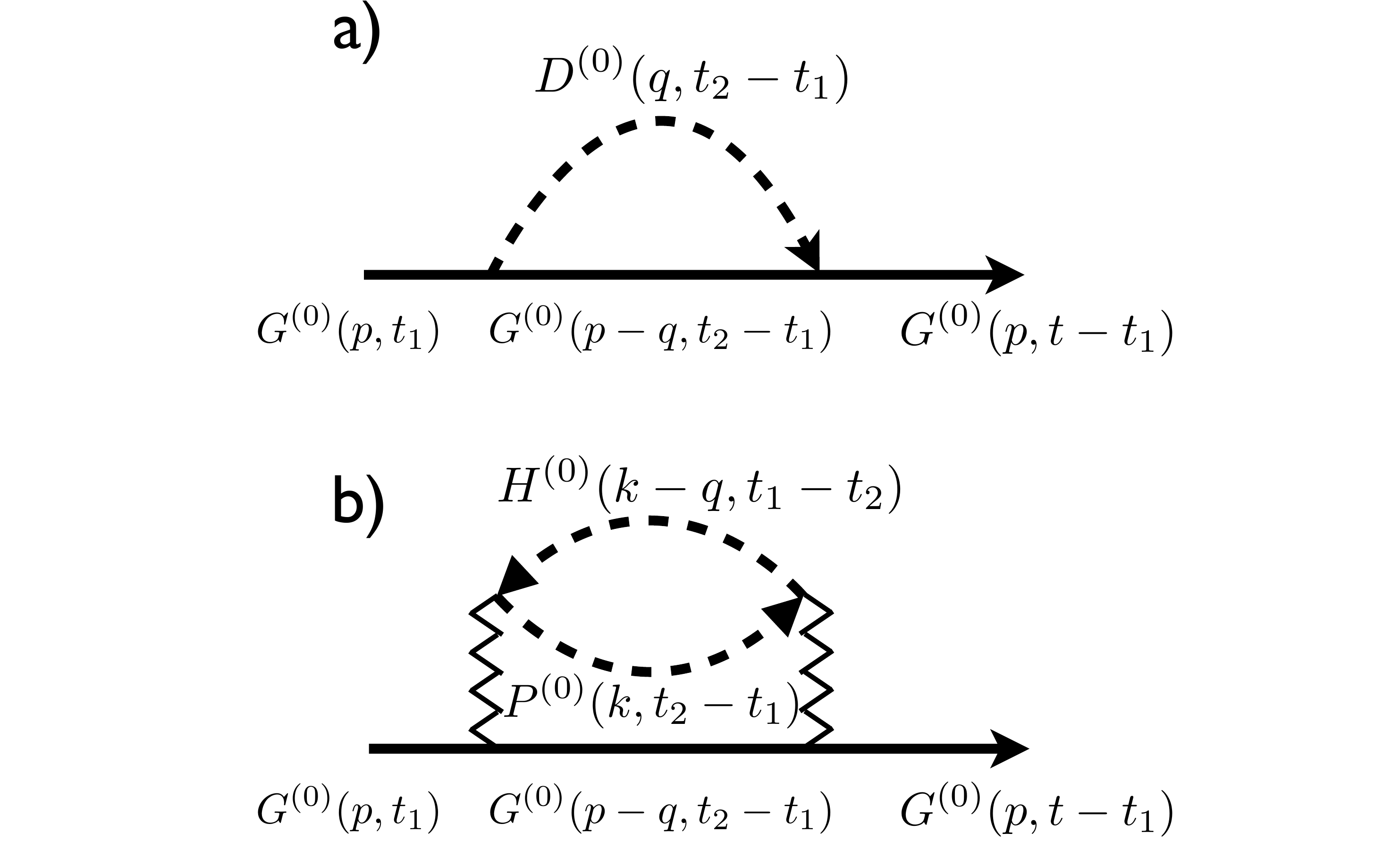}
\end{center}
\caption{a) Diagrammatic depiction of $W_2(p,t)$ for the linearized spectrum of a generic bath, where $G^{(0)}(p,t)=-i\theta(t)e^{-i\varepsilon_pt}$
and $D^{(0)}(p,t)=-i\theta(-t)e^{iv|q|t}-i\theta(t)e^{-iv|q|t}$ are the time-ordered Greens functions
of non-interacting impurity and bath respectively. b) Diagrammatic depiction of $W_2(p,t)$ for a bath
of free fermions, with $G^{(0)}(p,t)$ as in a), and $P^{(0)}(k,t)=-i(1-\theta(k_f-k))e^{i2\Ju\cos(k)t}$, 
$H^{(0)}(k,t)=i\theta(k_f-k)e^{i2\Ju\cos(k)t}$ the particle and hole propagators of free spinless fermions
respectively.}
\label{fig:Fig1_SM}
\end{figure}
The Linked Cluster Expansion (LCE) method represents
a different way to sum up the infinitely many Feynman diagrams
that come up in a perturbative expansion in the interaction
(the Dyson equation is the more well known). It has proven itself
able to capture the occurrence of the Orthogonality Catastrophe
for the immobile impurity in a Fermi liquid. It proceeds from the assumption
that the Greens function can be written as a re-exponentiated sum
\begin{equation}\label{def:lce}G(p,t)=G^{(0)}(p,t)e^{\sum_{n=1}^{\infty}F_n(p,t)},\end{equation}
where $G^{(0)}(p,t)=-i\theta(t)e^{-i\epsilon_pt}$ is the Greens function of the non-interacting
impurity, and $F_n(p,t)$ is the sum of all diagrams of $n$th order in the coupling $\Uud$.
They are determined from matching all terms of the same power in $\Uud$ in
the Taylor expansion of (\ref{def:lce}) to the standard perturbative expansion of $G(p,t)$ 
in the interaction picture (c.f.~\cite{BookMahan2000} for details).
\subsection{General quantum liquid in TLL approximation}\label{subsub_tll}
To employ LCE for the general Hamiltonian (3) of the article, we approximate $\OP{H}{}\approx \OP{H}{\rm LL} + \OP{H}{\rm imp} +\OP{H}{\rm coup}$.
Here, $\OP{H}{\rm LL} = \sum_q v |q| \hat{c}^{\dagger}_q\hat{c}_q$ describes the linearized 
single particle excitations of the $\up$-sector of $H$, represented by new bosonic operators $c_q$, $c_q^{\dagger}$,
with $v$ the sound velocity~\cite{BookGiamarchi2003}. The bath-impurity coupling can be approximated as
$\OP{H}{\rm coup}=\sum_{k,q} V(q) b^{\dagger}_{k+q\dn} b_{k\dn}(c_{q}+c^{\dagger}_{-q})$, with
$V(q)=\Uud\sqrt{\frac{K|q|}{2\pi L}}e^{-|q|/2q_c}$, where $K$ is the Luttinger
liquid parameter of $H_{\rm LL}$, and $q_c$ represents a momentum cutoff. The impurity Hamiltonian
can then be written as $ \OP{H}{\rm imp}=\sum_q \varepsilon_q b^{\dagger}_{q\dn}b_{q\dn}$.

Comparison of the Taylor expansion of (\ref{def:lce}) with the standard-perturbation expansion
yields that the first possible non-zero term in the LCE, $F_2(p,t)$, is equal to $e^{i\varepsilon_pt}W_2(p,t)$,
with $W_2(p,t)=\frac{(-i)^2}{2!}\int_0^t dt_1\int_0^t dt_2 \langle\mathcal{T}\OP{b}{p\dn}(t)\OP{H}{\rm coup}(t_1)\OP{H}{\rm coup}(t_2)\OP{b}{p\dn}^{\dagger}(0)\rangle$ 
the second-order contribution to the standard interaction-picture perturbation theory (diagram shown
in Fig. \ref{fig:Fig1_SM}a). A straightforward calculation yields
\begin{equation}\label{f2_tll}
F_2(p,t)=\int du \frac{1+itu-e^{itu}}{u^2}R(u),
\end{equation}
with $R(u)=\int dq V(q)^2\delta(u+\varepsilon_p-\varepsilon_{p+q}-v|q|)$, the density of excitations.

The behaviour of $R(u)$ and thus of $F_2(p,t)$ now depends critically on the $q$-roots inside the $\delta$-function.
For small impurity momenta, when the impurity dispersion $\varepsilon_p=-2\Jd\cos(p)$ can be approximated as $\varepsilon_p\approx\Jd p^2$,
all possible roots can be computed explicitly, 
\begin{eqnarray}
q_{\pm}^>&=&-\left(p+\frac{v}{2\Jd}\right)\pm\sqrt{\left(p+\frac{v}{2\Jd}\right)^2-\frac{u}{\Jd}}\\
q_{\pm}^<&=&-\left(p-\frac{v}{2\Jd}\right)\pm\sqrt{\left(p-\frac{v}{2\Jd}\right)^2-\frac{u}{\Jd}}
\end{eqnarray} 
of which only two, or none, may apply at a time (see eq. (\ref{ru_exp})).
Depending on whether $p<\frac{v}{2\Jd}$ or $p>\frac{v}{2\Jd}$ - corresponding to whether $\delta\epsilon(q)=v|q|$ has solutions only at $q=0$
or at $q\neq0$ as well - and the sign of $u$, $R(u)$ has several possible 
cases:
\begin{widetext}
\begin{equation}\label{ru_exp}
R(u) =
\begin{cases}
0 & p<\frac{v}{2\Jd} \wedge u > 0, \quad\mbox{roots: none}\\
\frac{1}{2\Jd}\left[\left(\frac{p-\frac{v}{2\Jd}}{\sqrt{\left(p-\frac{v}{2\Jd}\right)^2-\frac{u}{\Jd}}}-1\right)e^{-|q^<_+|/q_c}
+\left(\frac{p-\frac{v}{2\Jd}}{\sqrt{\left(p-\frac{v}{2\Jd}\right)^2-\frac{u}{\Jd}}}+1\right)e^{-|q^<_-|/q_c}\right] & p>\frac{v}{2\Jd} \wedge u > 0, \quad\mbox{roots: }q_{\pm}^< \\
\frac{1}{2\Jd}\left[\left(1-\frac{p+\frac{v}{2\Jd}}{\sqrt{\left(p+\frac{v}{2\Jd}\right)^2-\frac{u}{\Jd}}}\right)e^{-|q^>_+|/q_c}
+\left(1+\frac{p-\frac{v}{2\Jd}}{\sqrt{\left(p-\frac{v}{2\Jd}\right)^2-\frac{u}{\Jd}}}\right)e^{-|q^<_-|/q_c}\right] & u<0, \quad\mbox{roots: }q_{+}^>, q_{-}^<\\
\end{cases}
\end{equation}
\end{widetext}
As we are interested in the long-time behaviour of $\operatorname{Re}F_2(p,t)$, the integration over $u$ will be dominated by the small values around zero.
For $p<\frac{v}{2\Jd}$, it is straightforward to see that $R(u)\propto u$ for $0\leq -u\ll\Jd\left(p-\frac{v}{2\Jd}\right)^2$,
leading to $\operatorname{Re}F_2(p,t)\approx\int_0^{\infty} du \frac{1-\cos(tu)}{u}\propto\log(t)$, and thereby obtain ID behaviour. 
Conversely, when $p>\frac{v}{2\Jd}$, $R(u)$ is a finite constant around $u=0$, and therefore
$\operatorname{Re}F_2(p,t)\approx\int_0^{\infty} du \frac{1-\cos(tu)}{u^2}\propto t$, thus signaling the re-onset of quasiparticle
dynamics.

In the limit of small momenta and neglecting the dependency on the momentum cutoff, we can further compute the LCE-approximation
to $\Delta(p)$. Combining eqs. (\ref{f2_tll}) and (\ref{ru_exp}), one obtains 
$\operatorname{Re}[F_2(p,t)]\approx \frac{K\Uud^2}{2\pi^2v^2}\left(1+\frac{6\Jd^2}{v^2} p^2\right)\log(t)$
and thus $\Delta_{\rm LCE}(p)=-1+\frac{K\Uud^2}{2\pi^2v^2}\left(1+\frac{6\Jd^2}{v^2} p^2\right)$.

\subsection{Free fermions in 1D}
When the bath of $\up$-particles is given by free fermions, no approximation
of Hamiltonian (3) (see article) is necessary to perform the LCE. Matching the Taylor expansion of (\ref{def:lce})
to the standard perturbation theory yields two non-zero contributions, 
$F_1(p,t)=e^{i\varepsilon_p t}W_1(p,t)$, $F_2(p,t)=e^{i\varepsilon_p t}W_2(p,t)-\frac{1}{2!}F_1(p,t)^2$,
where $W_1(p,t)=-i\int_0^t dt_1 \langle\mathcal{T}\OP{b}{p\dn}(t)\OP{H}{\rm coup}(t_1)\OP{b}{p\dn}^{\dagger}(0)\rangle$,
$W_2(p,t)$ as above, and $\OP{H}{\rm coup}=\sum_{i}\Uud\hat{n}_{i\up}\hat{n}_{i\dn}$.

The only effect of $F_1(p,t)$ is to effect a Hartree shift of $\Uud\Nup/L$ in the frequency of $G(p,t)$, and further
to remove the disconnected diagram contained in $W_2(p,t)$. The only remaining diagram is shown in Fig. \ref{fig:Fig1_SM}b.
Calculating it, $F_2(p,t)$ is of the same form as in eq. (\ref{f2_tll}), with $R(u)$ given by eq. (\ref{lce2}).

\section{Extracting $\Delta(p)$ from DMRG: an example}
\begin{figure}[t]
\begin{center}
\vspace{-0cm}
\includegraphics[width=1\columnwidth,trim =100mm 32mm 100mm 30mm, clip]{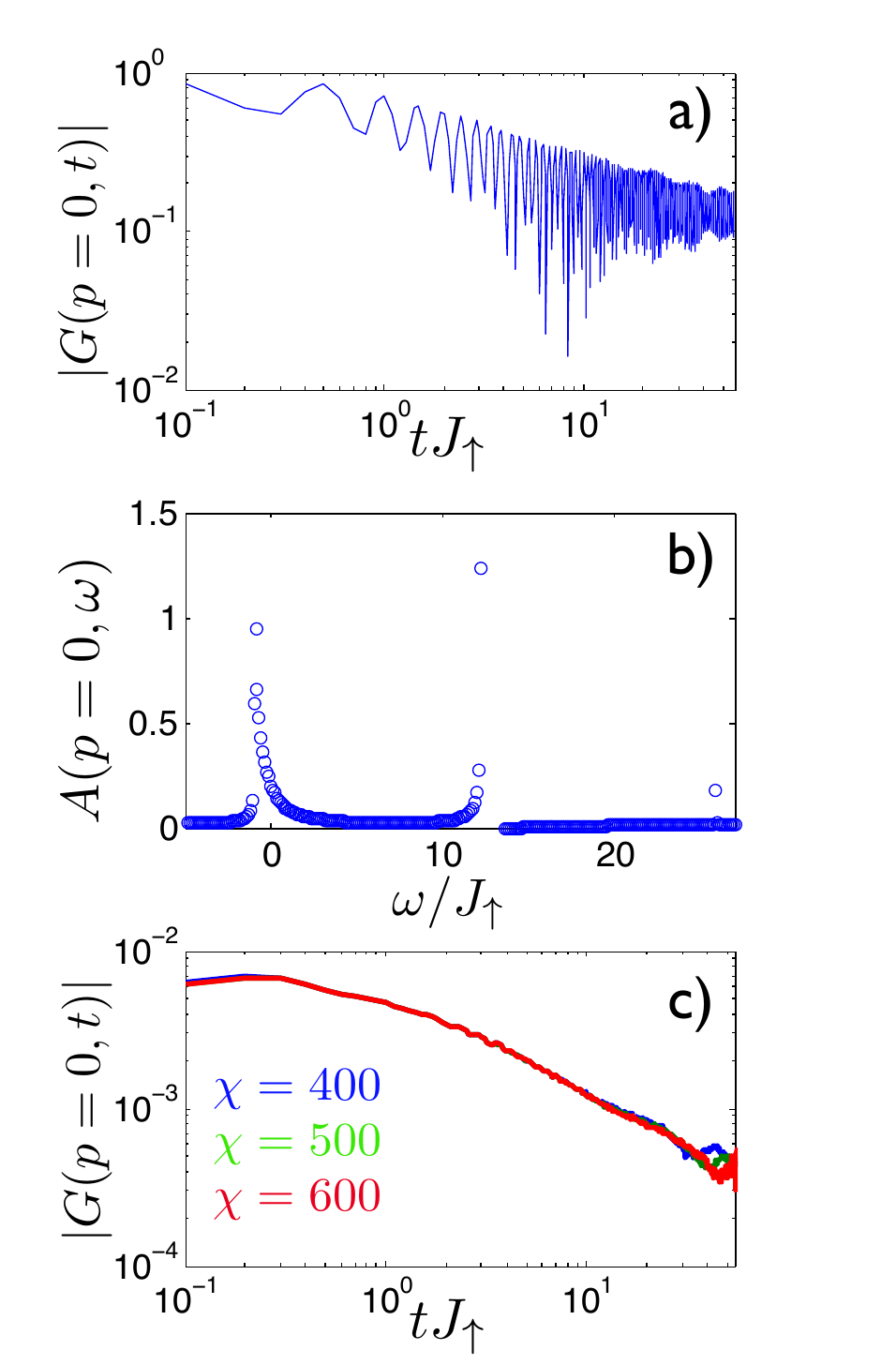}
\end{center}
\caption{Example of data analysis a) Raw, unfiltered numerical data for $|G(p=0,t)|$, for a system of softcore bosons, with $\Uu=2\Ju$,
$\Jd=\Ju$, $\Uud=10\Ju$, $L=201$, $\Nup=100$ and Schmidt-number $\chi=600$. b) Spectral function $A(p=0,\omega)$
obtained from a) via discrete Fourier transformation. Besides the expected power-law behaviour above threshold at the left,
two more such thresholds can be observed, corresponding to lattice-specific repulsively-bound impurity states (see text). c) $|G(p=0,t)|$ obtained from b) after Fourier transforming back to the time domain with a spectral filter, keeping only $\omega\in[-\Uud,\Uud]$, for three different $\chi$, i.e. accuracies. The apparent rising of $|G(p=0,t)|$ at long times is thus revealed as an artifact of entanglement increasing exponentially with the simulation time (see main text).}
\label{fig:Fig2_SM}
\end{figure}

To illustrate the procedure by which we obtain $\Delta(p)$ from DMRG simulations, we show here a step-by-step
example. This analysis starts from the insight that a mobile impurity in the infrared-dominated regime
will exhibit a Greens function behaviour $|G(p,t)| \propto t^{-1-\Delta(p)}$. We pick the case of a softcore boson bath ($\Uu=2\Ju$) with $\Ju=\Jd$, $\Uud=10\Ju$, $\Nup=100$ and $L=201$, at momentum $p=0$, as one that is generally representative for the required 
analysis of the raw numerical results. The number of Schmidt-components retained, here denoted as $\chi$, the critical quantity determining the accuracy of any static or time-dependent DMRG, will depend on the required accuracy (see below). Ultimately, the final result considered in this example will be at $\chi=600$

Shown in Fig.~\ref{fig:Fig2_SM}a is $|G(p=0,t)|$, computed directly from the raw output
$G(x,t)=e^{iE_{GS}t}\langle GS|b_{(L+1)/2-x\dn}e^{-iHt}b^{\dg}_{(L+1)/2\dn}|GS\rangle$ via discrete Fourier transformation 
$G(p,t)=\sum_{x=-(L-1)/2}^{(L-1)/2}e^{-ipx}G(x,t)$. As described in the main text, this raw output is computed by propagating 
the state $b^{\dg}_{(L+1)/2\dn}|GS\rangle$ in time, and computing all overlaps with the states 
$b^{\dg}_{(L+1)/2-x\dn}|GS\rangle$ after every time step. In Fig. ~\ref{fig:Fig2_SM}b, we show the resulting spectral function, $A(p=0,\omega)=-\frac{1}{\pi} \operatorname{Im}[G(p=0,\omega)]$, with $G(p,\omega)=\sum_{t_n=0}^{T_f}e^{-i\omega t_n}G(p,t_n)$ being the Fourier transform to the frequency domain and $t_n=n\Delta t$ being discretized time. As usual for the discrete Fourier transform, the available resolution
in $\omega$ is given by $2\pi/T_f$. 

The structure of $A(p=0,\omega)$ is revealing: besides the expected 
power-law threshold at about $\omega_0=-0.98$, we observe two more divergences, one at $\omega_1=12.25$,
and another very weak one at $\omega_2=25.86$. Such higher-frequency divergences are a general feature when $\Uud\geq 2\Ju+2\Jd$,
and they are due to the impurity tunneling only between sites carrying one $\up$-particle each ($\omega_1$)
or two each ($\omega_2$). These divergences are thus still due to ID-dynamics of an impurity, but one that is repulsively bound~\cite{Winkler2006} to one $\up$-particle or two $\up$-particles respectively. We are currently not interested in these lattice-specific high-energy features
of the impurity, even though they also show ID behaviour, albeit governed by different exponents. We want to focus on the low-energy properties of the impurity, the ones that are expected to align quantitatively with the ones from those baths in the continuum that have the same TLL-parameters as the lattice-confined bath. 

For this, we spectrally filter $G(p,\omega)$, keeping only frequencies $\omega\in[-\Uud,\Uud]$, whenever additional high-energy divergences from repulsively bound impurity-states appear. As obtaining $\Delta(p)$ from fitting directly to $A(p,\omega)$ carries
much too large an error due to the limited $\omega$ resolution, we Fourier-transform the spectrally filtered $G(p,\omega)$ back into the
time domain.  The absolute value of this Greens function is shown in Fig.~\ref{fig:Fig2_SM}c for three different values of the Schmidt-number $\chi$. After the initial, non-universal short time regime, we observe a clear power-law decay in time, albeit with a slope that is different from that of the unfiltered $G(p,t)$ in Fig.~\ref{fig:Fig2_SM}a, due to the power law divergences at $\omega_{1,2}$ having been filtered out. Eventually, at long times the curve starts bending up again, which however is only due to the well known limitation of all time-dependent DMRG for out-of-equilibrium systems, the exponential growth of the entanglement that needs to be retained in order to describe them accurately~\cite{Schollwock2011}. As a consequence, we see the up-bending pushed systematically out in time with increased $\chi$. Eventually, once the excitations of the bath reach the open boundaries of the finite system, another up-bending will occur (not shown here), which is a real effect, and not an artifact due to the loss of accuracy at long times.

Finally, from the slope of $|G(p=0,t)|$ in Fig.~\ref{fig:Fig2_SM}c a simple least-square fit allows accurate extraction of $\Delta(p=0)$.

\section{Persistence of ID dynamics at small $\Jd$ for a bath of softcore bosons}
\begin{figure}[t]
\begin{center}
\vspace{-0cm}
\includegraphics[width=1\columnwidth,trim =60mm 68mm 70mm 68mm, clip]{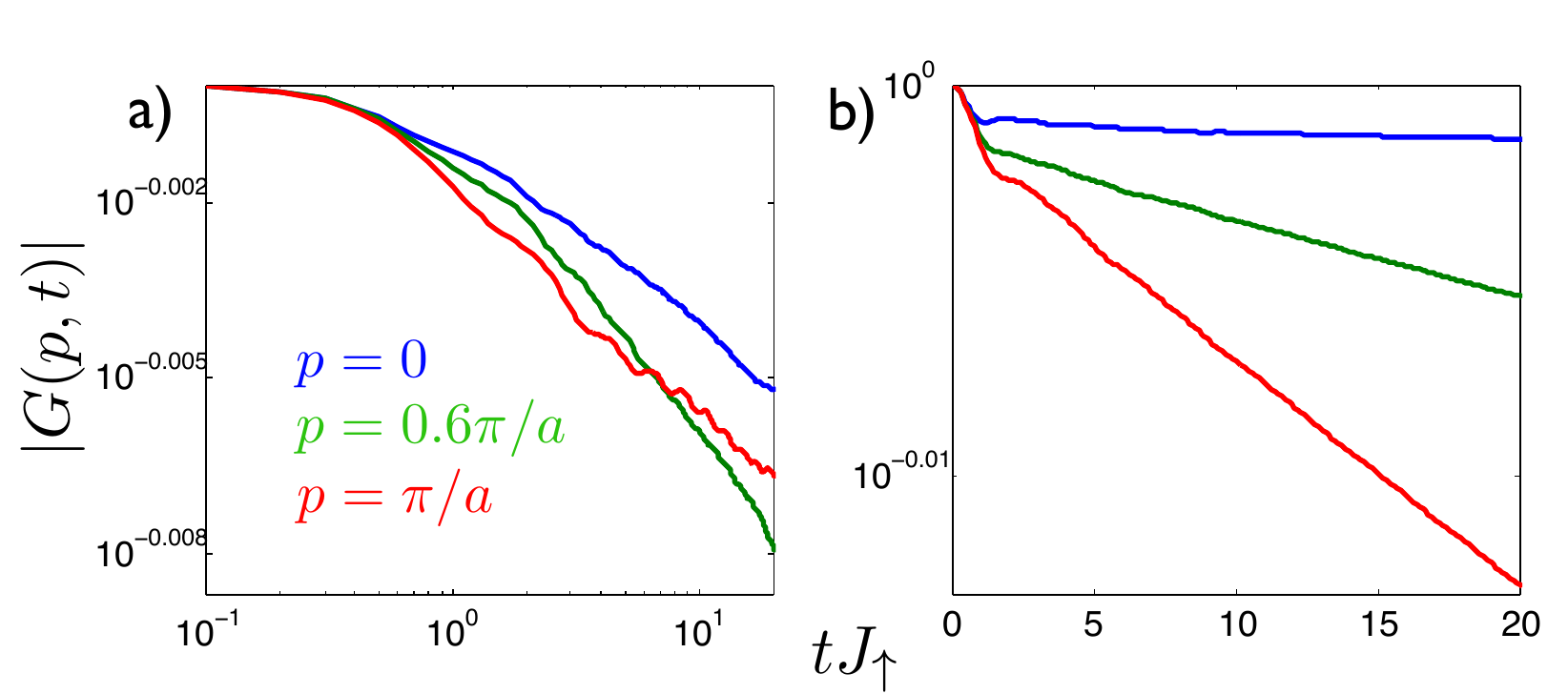}
\end{center}
\caption{Comparing the persistence of impurity ID dynamics to all quasimomenta in softcore bosons with the transition
to QP-behaviour in a bath of free fermions, at $\Jd=0.2\Ju$, $\Uud=0.2\Ju$, $\Nup=100$ and $L=201$. a) The bath is comprised of softcore bosons, $\Uu=2\Ju$,  $|G(p,t)|$ is shown for three momenta throughout the Brilloin-zone on a log-log scale. We observe power-law behaviour throughout, including the Brilloiun-zone edge $p=\pi/a$. Thus, impurity-dynamics is always ID. b) For comparison purposes, $|G(p,t)|$ is plotted at the same momenta semi-logarithmically for a bath of free fermions. As always in this case, once $p>k_F$ $|G(p,t)|$ decays exponentially, signaling the reemergence of the QP regime.}
\label{fig:Fig3_SM}
\end{figure}
The striking possibility that ID dynamics may persist at any quasimomentum 
strongly distinguishes a bath of softcore bosons from one of free fermions ($\Uu=\infty$).
The effect is due to the fact that the SPES of softcore bosons, shown in Fig. 1c of the main text for
a 1D continuum model, have nonvanishing weight only for small momenta. Thus, when $\delta\epsilon(q)$
has a small amplitude due to low $\Jd$, it is possible for the impurity never to emit real, finite-wavelength
phonons into the bath due to $\delta\epsilon(q)$ never intersecting with the SPES at appreciable weight.
As explained in the main text, for free fermions (and sufficiently repulsive bosons s.t. their SPES is close to that of free fermions),
this effect is strictly precluded due to the fact that their SPES always bends down to zero energy with finite weight
at $q=\pm 2k_F$.
Here, we illustrate this very different behaviour by showing $|G(p,t)|$ in Fig.~\ref{fig:Fig3_SM} at three different quasimomenta,
$p=0$, $=0.6\pi/a$ and $=\pi/a$, in subfigure a) for softcore
bosons with $\Uu=2\Ju$ on log-log scale, in b) for a bath of free fermions on log-scale. In both cases we have $\Jd=0.2\Ju$, $\Uud=0.2\Ju$, $\Nup=100$ and $L=201$.
As is readily visible, the impurity in the softcore boson bath retains its ID-dynamics above $k_F=\pi n_{\up}/a$, exhibiting
a power-law behaviour $|G(p,t)|\propto t^{-1-\Delta(p)}$ throughout,
while in the fermion bath it decays exponentially and behaves as a quasiparticle again, just the same as shown
in Fig. 1a and b of the main text for $\Jd=\Ju$.

\section{Deriving $f$ as a function of the correlation hole size}
To relate $f=\lim_{\Uud\rightarrow\infty}2(\Delta(0)+1)$ to a static property of the ground state
of the bath \textit{in presence} of the impurity, we start with the observation
that both Bethe-ansatz solutions for lattice fermions as well as exact diagonalisation
solutions for small systems of arbitrary parameters show that for $\Uud\rightarrow\infty$
$\Delta(0)$ (as given by eq. (6)) is proportional to $E_{\rm GS}(\Nup=N,\Ndn=1)$
To make this relationship quantitative, it is easiest to commence from a general two-component
Lieb-Lieninger (i.e. first-quantized) Hamiltonian for impurity and bath, and use a result obtained by 
Lamacraft~\cite{Lamacraft2009}, namely that in the limit of a nearly opaque impurity in a 1D bath
$\Delta(0)=-1+\frac{K}{2}\left(\frac{\uphi}{\pi v}\right)^2$, where $K$ and $v$ are the Luttinger liquid
parameters of the baths' low-energy effective theory (see above) and $\uphi$ is the effective
forward scattering amplitude of the impurity (see below).

In the lab frame th,e calculation thus begins with the first-quantized Hamiltonian
\begin{equation}\label{h_lab_frame}
\OP{H}{}=\sum_{i=1}^{\Nup}\frac{\hat{p}_i^2}{2m}+\sum_{i<j}\Uu(\hat{x}_i-\hat{x}_j)+\frac{\hat{P}^2}{2M}+\sum_{i=1}^{\Nup}\Uud(\hat{x}_i-\hat{X})\end{equation}
where $\OP{x}{i}$, $\OP{X}{ }$ ($\OP{p}{i}$, $\OP{P}{i}$) denote position (momentum) operators of $\uparrow$-particles of mass $m$ and
$\downarrow$-impurity of mass $M$ respectively. The interactions inside a generic bath $\Uu$ will depend only on $\OP{x}{i}-\OP{x}{j}$,
while the bath-impurity interaction will depend on $\OP{x}{i}-\OP{X}{ }$. Boosted to the impurity rest-frame
through the unitary $U=\exp(-i\OP{X}{}\sum_{i=1}^{\Nup}\hat{p}_i)$, the Hamiltonian reads:
\begin{eqnarray}\label{h_imp_frame}
\OP{H}{}' & = & \OP{U}{}^{\dagger}\OP{H}{}\OP{U}{} = \sum_{i=1}^{\Nup}\frac{\hat{p}_i^2}{2m}+\sum_{i<j}\Uu(\hat{x}_i-\hat{x}_j)\nonumber\\
&&+\frac{\left(\hat{P}-\sum_{i=1}^{\Nup}\hat{p}_i\right)^2}{2M}+\sum_{i=1}^{\Nup}\Uud(\hat{x}_i).\end{eqnarray}

If $|GS\rangle$ denotes the ground state of Hamiltonian (\ref{h_lab_frame}) and $\OP{\rho}{}(y)=\sum_{i=1}^{\Nup}\delta(y-\OP{x}{i})$ 
the baths local density operator, it follows directly
\begin{equation}\label{op_eq}
\langle GS|\OP{U}{}^{\dagger} \OP{\rho}{}(y)\OP{U}{} |GS\rangle = \sum_{i=1}^{\Nup}\langle GS| \delta(y-\OP{x}{i}+\OP{X}{}) |GS\rangle.
\end{equation}
In second quantization, $\sum_{i=1}^{\Nup} \delta(y-\OP{x}{i}+\OP{X}{})$ translates to $\sum_{x,X} \delta(y-x+X)\OP{n}{x\up}\OP{n}{X\dn}$

The strategy to obtain eq. (7) of the article, is to compute the left-hand side of eq. (\ref{op_eq}) in the impurity rest-frame
using the low-energy TLL-approximation of Hamiltonian (\ref{h_imp_frame}) (c.f.~\cite{BookGiamarchi2003,Lamacraft2009}),
where e.g. the baths density operator is approximated as $\OP{\rho}{}(y)\approx \frac{\Nup}{L}+\frac{1}{\pi}\nabla\OP{\phi}{}(y)$,
and the $\OP{H}{}'$ as
\begin{eqnarray}\label{h_imp_frame_tll}
\OP{H}{}' & \approx & \OP{U}{}^{\dagger}\OP{H}{}\OP{U}{} = \frac{v}{2\pi}\int dx K(\nabla\OP{\theta}{}(x))^2+K^{-1}(\nabla\OP{\phi}{}(x))^2\nonumber\\
&&+\frac{1}{2M}\left(\OP{P}{}-\frac{1}{\pi}\int dx \nabla\OP{\theta}{}(x)\nabla\OP{\phi}{}(x)\right)^2\nonumber\\
&&+\frac{\uphi}{\pi}\int dx \lambda(x)\nabla\OP{\phi}{}(x).
\end{eqnarray}
Here, $\OP{\phi}{}(x)$ and $\OP{\theta}{}(x)$ represent the standard conjugate fields for the density and phase fluctuations of the bath,
and $\lambda(x)$ is some positive function narrowly peaked around $x=0$, s.t. $\int_{-d}^d dx \lambda(x) =1$ (how $a$ is set is discussed
below).

With this approximate Hamiltonian, we can calculate
\begin{equation}
\langle GS|\OP{U}{}^{\dagger} \OP{\rho}{}(y)\OP{U}{} |GS\rangle \approx \frac{\Nup}{L}+\frac{1}{\pi}\langle\nabla\OP{\phi}{}(y)\rangle_{\OP{H}{}' }
\end{equation}
where, using the path integral representation,
\begin{widetext}
\begin{equation}
\langle\nabla\OP{\phi}{}(y)\rangle_{\OP{H}{}' }=\lim_{\beta\rightarrow\infty}\frac{\int[\mathcal{D}\phi(x,\tau)][\mathcal{D}\theta(x,\tau)][\mathcal{D}P(\tau)]\nabla\phi(y,0) e^{S[\phi(x,\tau),\theta(x,\tau),P(\tau)]}}
{\int[\mathcal{D}\phi(x,\tau)][\mathcal{D}\theta(x,\tau)][\mathcal{D}P(\tau)]e^{S[\phi(x,\tau),\theta(x,\tau),P(\tau)]}}
\end{equation}
\end{widetext}
where the action $S[\phi(x,\tau),\theta(x,\tau),P(\tau)]$ is given by
\begin{equation}
S=\int_0^{\beta}d\tau\int dx i\theta(x,\tau)\partial_{\tau}\phi(x,\tau) -H'[\phi(x,\tau),\theta(x,\tau),P(\tau)]
\end{equation}
Completing the square for $\nabla\phi(x,\tau)$ in the action and defining a new field
$\phi_s(x,\tau)=\phi(x,\tau)+\frac{K\uphi}{v}\int_{-\infty}^x dx'\lambda(x')$
yields an action
\begin{widetext}
\begin{eqnarray}\label{eff_action}
S&=&\int_0^{\beta}d\tau\left[\int dx i\theta(x,\tau)\partial_{\tau}\phi_s(x,\tau) - 
\frac{v}{2\pi}\left(\int dx K(\nabla\theta(x,\tau))^2+K^{-1}(\nabla\phi_s(x,\tau))^2\right)\right.\nonumber\\
&&\left.-\frac{1}{2M}\left(P(\tau)-\frac{1}{\pi}\int dx \nabla\theta(x)\left(\nabla\phi_s(x,\tau)-\frac{K\uphi}{v}\lambda(x)\right)\right)^2
+\int dx\frac{(K\uphi\lambda(x))^2}{2v\pi}
\right].
\end{eqnarray}
\end{widetext}
We can assume that the term $\int dx \nabla\theta(x)\lambda(x)$ is effectively
zero, as there will be no current across the position of the nearly opaque impurity
(c.f.~\cite{Lamacraft2009}). With this, the action (\ref{eff_action}) becomes one where
bath and impurity are decoupled (the final term, the counterterm, is obviously irrelevant
for any observable), thus $\langle\nabla\phi_s(y,0)\rangle_S=0$, and we obtain
\begin{equation}
\langle GS|\OP{U}{}^{\dagger} \OP{\rho}{}(y)\OP{U}{} |GS\rangle \approx \frac{\Nup}{L}-\frac{K\uphi}{v\pi}\lambda(y)
\end{equation}
Together with Lamacrafts' result on $\Delta(0)$ for the opaque impurity, integrating from $-d$ to $d$ over the position $y=0$ in the impurity
rest frame in eq. (\ref{op_eq}) then immediately yields the main result, eq. (7) of the article. As to how $d$ is chosen: the Tomonaga-Luttinger liquid approximation is designed to capture low-energy, long-wavelength physics. Thus, $d$ has to be at least large enough s.t. the impurity-caused density depletion of the bath deviates only weakly from its' background density. As long as it satisfies this requirement, we expect that due to the structure of eq. (7) in the article, the resultant $f$ will not depend very much on the specific value of $d$, as $\frac{\Nup}{L^2}$ (the background value of $\langle\hat{n}_{x\up}\hat{n}_{0\dn}\rangle$) is already subtracted from it.

\end{document}